\documentclass[11pt,a4paper]{article}
\usepackage{float}
\usepackage{jheppub} % for details on the use of the package, please
\def\Z#1{\mathbb{Z}_{#1}}
\def\odp{\mathsf{D}}
\def\tr{\mathsf{T}}
\def\nn{\nonumber\\ }

\newcommand{\D}{\mathcal{D}}
\renewcommand{\O}{\mathcal{O}}

\begin{document}

%%%%%%%%%%%%%%%%%%%%%%%%%%%%%%%%%%%%%%%%%%%%%%%%%%%%%
\title{Operator Dimension Parity Fractionalization}

\author{Christopher W.~Murphy}
\emailAdd{chrismurphybnl@gmail.com}

\abstract{
Lorentz invariant quantum field theories (QFTs) with fermions in four spacetime dimensions (4D) have a $\Z4$ symmetry provided there exists a basis of operators in the QFT where all operators have even operator dimension, $d$, including those with $d > 4$.
The $\Z4$ symmetry is the extension of operator dimension parity by fermion number parity.
If the $\Z4$ is anomaly-free, such QFTs can be related to 3D topological superconductors. 
Additionally, imposing the $\Z4$ symmetry on the Standard Model effective field theory severely restricts the allowed processes that violate baryon and lepton numbers.
}
\maketitle

%%%%%%%%%%%%%%%%%%%%%%%%%%%%%%%%%%%%%%%%%%%%%%%%%%%%%
\section{Introduction}

Qualitatively new phenomena can emerge given a sufficiently wide separation of scales~\cite{Anderson:1972pca}.
However, the same theoretical tools can describe seemingly unrelated systems.
A prominent example is the superconductor (SC) of condensed matter physics and the Standard Model (SM) of particle physics.
They have vastly different scales.
A conventional SC typically has a critical temperature around $T_c \sim 10~\rm{K} \sim 10^{-12}~\rm{GeV}$. 
In contrast, the electroweak (EW) vacuum expectation value of the SM is $v \sim 10^2~\rm{GeV}$.
Nevertheless, both Ginzburg-Landau theory~\cite{Ginzburg:1950sr} and the Higgs mechanism~\cite{Englert:1964et, Higgs:1964pj, Guralnik:1964eu} of the SM describe a spontaneously broken continuous symmetry where gauge bosons receive mass from would-be Nambu-Goldstone bosons~\cite{Nambu:1960tm, Goldstone:1961eq, Goldstone:1962es}.
This example can be extended because the Higgs mechanism is present in more quantum field theories (QFTs) than just the SM.

In a contemporary twist to the story, there is a relationship between a topological superconductor (TSC) in three spacetime dimensions (3D) and the SM augmented with sterile neutrinos ($\nu$SM) in 4D~\cite{You:2014oaa, Tachikawa:2018njr, Garcia-Etxebarria:2018ajm}.
Both theories have a discrete $\Z4$ symmetry with an anomaly that requires the number of fermions (of the same handedness) to be $0 \bmod 16$~\cite{Garcia-Etxebarria:2018ajm, Hsieh:2018ifc, Witten:2015aba}.
Although our general result, to be discussed, will show that both the SM and $\nu$SM have a candidate $\Z4$ symmetry, the increase in the number of fermions per generation from 15 to 16 is what allows $\nu$SM to be related to a TSC in an anomaly-free way.

The $\Z4$ symmetry of the TSC is time reversal~\cite{Witten:2015aba}, raising the question of: what is the interpretation of the $\Z4$ in the 4D theory?
Additionally, given that extensions of the original connection between SCs and the SM are possible, it is natural to ask: is the relation involving TSCs specific to $\nu$SM, or is there a larger class of 4D QFTs where this is possible?
A hint to the second question's answer comes from~\cite{Ma:2023yxq, Ma:2023fin}, which recently noted that renormalizable theories with only dimension-2 and -4 operators have a $\Z4$ symmetry.

In this work, we give a unified answer to both questions.
Our critical insight is to consider Lorentz invariant 4D QFTs with fermions in including those with operators with canonical dimension $d > 4$, that is effective field theories (EFTs).
We show that if a basis exists in such a QFT where all the operators have even canonical dimension, then the QFT possesses a $\Z4$ symmetry.
This immediately suggests an interpretation of the symmetry as operator dimension parity, $\odp$.
We go on to discuss aspects of the symmetry, including why it is a $\Z4$ rather than $\Z2$ symmetry, the full symmetry group including spacetime symmetries, as well as the anomaly-free conditions needed to determine whether there is an obstruction to gauging the global symmetry in the quantum theory.
We also show any irreducible representation (irrep) of the Lorentz group in a given power counting scheme, that need not be mass dimension, has a well-defined $\odp$ charge.
Finally, compelled by the central role of sterile neutrinos in the relationship to TSCs, we investigate the consequences of imposing the $\Z4^{\odp}$ symmetry on the Standard Model EFT with sterile neutrinos ($\nu$SMEFT), finding that the symmetry severely restricts the types of allowed baryon and lepton number violating processes while still allowing for baryon minus lepton number violation.

%%%%%%%%%%%%%%%%%%%%%%%%%%%%%%%%%%%%%%%%%%%%%%%%%%%%%
\section{$\Z4$ Operator Dimension Parity}

If a Lagrangian only contains operators, $\O$, of even canonical dimension, $d \in 2\Z{+}$, then there is clearly a $\Z2$ symmetry, $\Z2 \O = (-1)^d \O = \O$, which we call operator dimension parity.
The Lorentz invariant operators, which are bosons, are in a linear representation of the $\Z2$ symmetry.
If the field content of the theory is also purely bosonic, e.g. a real scalar with a $\phi^4$ interaction, then this is the end of the story.
On the other hand, theories containing fermionic fields are the more interesting case as fermions sit in a projective representation of this $\Z2$ symmetry.
We want to find the central extension of the $\Z2$ where fermions are in a linear representation.

To get started, consider a Lorentz invariant QFT in $3 + 1$ spacetime dimensions with scalars, $\phi$, left- and right-handed Weyl fermions, $\psi$ and $\psi^\dagger$, and field strengths, $X_{\mu\nu}$.
Here, we use mass dimension as our power counting scheme and will generalize to all irreps of the Lorentz group and other power counting schemes later.
A Lorentz invariant operator in the Lagrangian is a monomial of the fields and covariant derivative, $\D_\mu$, 
\begin{equation}
\label{eq:op}
\O \sim X_L^{n_{-1}} \psi^{n_{-1/2}} \phi^{n_0} \psi^{\dagger n_{+1/2}} X_R^{n_{+1}} \D^{n_\D}
\end{equation}
where the exponent of each field, $n_h$, is labeled by its helicity, $h = j_r - j_l$ with $(j_l, j_r)$ being its irreducible representation under the Lorentz group, $G = Spin(3, 1) \hookrightarrow Spin(4; \mathbf{C}) \cong SU(2)_l \times SU(2)_r$.
Derivatives transform as bi-fundamentals of the Lorentz group, and we decompose the field strength into its irreps under the Lorentz group, $X_{L, R}$.
A necessary condition for an operator to be Lorentz invariant is that it must have an even number of fundamental indices of both $SU(2)_l$ and $SU(2)_r$
\begin{equation}
\label{eq:lorentz}
2 n_{l,r} = 2 n_{\mp1} + n_{\mp1/2} + n_\D .
\end{equation}
In Eq.~\eqref{eq:lorentz} $n_{l,r}$ is the number of epsilon symbols needed to contract the indices of $SU(2)_{l,r}$.
From Eq.~\eqref{eq:lorentz}, we derive the following constraint imposed by Lorentz invariance
\begin{equation}
\label{eq:constraint}
n_{-1/2} \bmod 2 = n_{+1/2} \bmod 2 = n_\D \bmod 2,
\end{equation}
which states that the number of left- and right-handed fermions and the number of derivatives in an operator must either all be even or all be odd.
The mass dimension of the operator, $[ \O ] = d$, is the sum of the mass dimensions of its constituents
\begin{equation}
\label{eq:dmassdim}
d = 2 \left(n_{-1} + n_{+1}\right) + \frac{3}{2} \left(n_{-1/2} + n_{+1/2}\right) + n_0 + n_\D ,
\end{equation}
with $d \in \Z+$.
The leftmost term on the right-hand side of Eq.~\eqref{eq:dmassdim} is always even.
We can use~\eqref{eq:constraint} to rearrange Eq.~\eqref{eq:dmassdim} arriving at
\begin{equation}
\label{eq:result}
\frac{3}{2} \left(n_{-1/2} - n_{+1/2}\right) + n_0 \equiv d \pmod 2 .
\end{equation}
Consider $d$ even, multiply both sides of~\eqref{eq:result} by $\pi$, and we have group elements for $\Z4$.

Thus, even-dimensional operators are invariant under a $\Z4$ symmetry, operator dimension parity, which we denote by $\odp$.
The charges are integers modulo four
\begin{equation}
\label{eq:charges}
\odp_{X_{\mu\nu}} = 0,\quad \odp_{\psi^\dagger} = 1,\quad \odp_{\phi} = 2,\quad \odp_{\psi} = 3 ,
\end{equation}
and we are free to swap the charge assignments of the left- and right-handed fermions.
Applying these charges to the expression for an operator in~\eqref{eq:op}, we find
\begin{equation}
\label{eq:z4op}
\odp \O = (-1)^d \O, \quad \odp \psi = -i \psi 
\end{equation}
where now both fermions and bosons are in linear representations of the symmetry.
From Eq.\eqref{eq:z4op}, we see that a QFT is invariant under $\odp$ if a basis exists where all the operators have even mass dimension.\footnote{
	This does not necessarily mean no operators associated with the theory have odd dimensions. 
	For example, gauge theories have conserved currents, $J_\mu$, which are dimension-3 operators. 
	However, the Lorentz invariant way a current shows up in a Lagrangian is in conjunction with a vector gauge field, $A^\mu J_\mu$, \textit{i.e.} by forming an even operator.
}
The renormalizable Standard Model is a fascinating theory that falls into this class.

%%%%%%%%%%%%%%%%%%%%%%%%%%%%%%%%%%%%%%%%%%%%%%%%%%%%%
\section{Aspects of the Symmetry}

Acting twice with $\Z4^\odp$ is equivalent to acting once with fermion number parity, $\Z2^F$, i.e. $\odp^2 = (-1)^F$.
From this we see that $\Z4^\odp$ is the central extension of operator dimension parity by fermion number parity
\begin{equation}
1 \rightarrow \Z2^F \rightarrow \Z4^\odp \rightarrow \Z2^\odp \rightarrow 1.
\end{equation}
When fermions are in a projective representation of the original symmetry and a linear representation of the extended symmetry, it is said that there is symmetry \textit{fractionalization}, see e.g.~\cite{Wang:2021nmi}.
For example, one could say that the TSC has time-reversal symmetry fractionalization as the time-reversal symmetry there satisfies $\tr^2 = (-1)^F$~\cite{Witten:2015aba}, making it an extension of the usual time-reversal by fermion number parity.
Here there is operator dimension fractionalization in the literal sense in the presence of fermionic operators, as fermions have half-integer operator dimensions.

If the $\odp$ symmetry is broken, either explicitly or spontaneously, the fermion number parity subgroup is always preserved.
However, $\Z2^F$ is also an element of the Lorentz group, $G$.
In particular, it is a $2\pi$ rotation that flips the sign of all the fermions.
Therefore, the total symmetry group is a semidirect product of the Lorentz group and operator dimension parity
\begin{equation}
\label{eq:spinZ4}
Spin^{\Z4} = G \ltimes \Z4^\odp = \frac{G \times \Z4^\odp}{\Z2^F}
\end{equation}
where the notation $Spin^{\Z4}$ is from~\cite{Tachikawa:2018njr}.

We would like to know when $\Z4^\odp$ has an 't Hooft anomaly~\cite{tHooft:1979rat} that prevents the global symmetry in the quantum theory from being gauged.
Complications in determining discrete anomalies include that the anomaly depends on the normalization of the charges and the total symmetry group and that there are no one-loop triangle diagrams for discrete symmetries.
Fortunately, within the past decade, tremendous progress has been made, resulting in a unified approach to studying anomalies -- discrete or continuous, local or global -- see~\cite[\textsection3]{Cordova:2022ruw} for a brief review.
The original mathematical work was by Dai and Freed~\cite{Dai:1994kq}, and an influential physics paper on the topic is~\cite{Witten:2015aba}.
As a practical matter, we will quote the relevant results from~\cite{Garcia-Etxebarria:2018ajm, Hsieh:2018ifc, Guo:2018vij}.
For left- and right-handed Weyl fermions with conjugate charges under the symmetry group~\eqref{eq:spinZ4}, the anomaly-free condition is
\begin{equation}
\label{eq:anom16}
n_{-1/2} - n_{+1/2} \equiv 0 \pmod {16} .
\end{equation}
On the other hand, when the charge assignments are the same, but the symmetry group is instead $G \times \Z4$ (without a quotient), the anomaly-free condition is instead
\begin{equation}
\label{eq:anom}
n_{-1/2} - n_{+1/2} \equiv 0 \pmod 4 .
\end{equation}
Early results on discrete anomalies relying on bespoke methods such as embedding a discrete symmetry into an auxiliary continuous symmetry also find Eq.~\eqref{eq:anom} for a $\Z4$ symmetry, see \textit{e.g.}~\cite{Banks:1991xj, Ibanez:1992ji, Csaki:1997aw}.
We mention this because a discrete anomaly can be trivialized, for example, by embedding it in a continuous symmetry.
However, one then must check that the resulting continuous symmetry is anomaly-free.
For instance, if one adds a new $U(1)$ symmetry to the SM \textit{without} introducing additional fermions, the only charge assignments for the SM fermions under the new $U(1)$ that prevent an ABJ anomaly are, up to an overall constant, their hypercharge assignments.

A $\Z2$ symmetry, in contrast with $\Z4$, has no known anomalies~\cite{Banks:1991xj, Ibanez:1992ji, Csaki:1997aw, Hsieh:2018ifc}.
Therefore, if for some reason there is a basis of operators in a QFT that only have even canonical dimensions, there would be no obstruction to gauging $\Z2^\odp$ regardless of the status of $\Z4^\odp$.
An alternative way to view this situation is to use dimensional analysis to make a statement about the non-renormalization of operators under $\Z2^\odp$.
Dimensional analysis in 4D, see \textit{e.g.}~\cite{Gavela:2016bzc}, states that an operator of canonical dimension $[\O] = d$ can be renormalized by a set of operators, $\{j\}$, with $[\O_j] = d_j$ provided that
\begin{equation}
\label{eq:nonrenorm}
d - 4 = \sum_j \left(d_j - 4\right)
\end{equation}
Importantly, Eq.~\eqref{eq:nonrenorm} is true to all-orders in the loop expansion~\cite[\textsection5.3]{Manohar:2018aog}.
It is the statement that an operator of even(odd) canonical dimension cannot be renormalized by an odd(even) number of insertions of operators with odd dimension.
An implication is that the anomalous dimension of operators that violate the symmetry must be proportional to the Wilson coefficient of at least one operator that also violates the symmetry.
Ref.~\cite{Jenkins:2017dyc} computed the one-loop anomalous dimensions in the Low-Energy EFT Below the EW Scale (LEFT) up to terms of dimension-6, and the resulting anomalous dimension matrices are an explicit example of the constraint imposed by~\eqref{eq:nonrenorm}.

%%%%%%%%%%%%%%%%%%%%%%%%%%%%%%%%%%%%%%%%%%%%%%%%%%%%%
\section{Conformal Center Symmetry}

We now generalize the result characterized by Eqs.~\eqref{eq:result},~\eqref{eq:charges}, and~\eqref{eq:z4op} to all irreps of the Lorentz group.
In doing so, we also find that it applies to more power counting schemes than just mass dimension.
As our proof of the existence of the $\odp$ symmetry only relies on Lorentz invariance and dimensional analysis, our approach is to try to combine those two factors.
A transformation under the conformal group, $Spin(4, 2)$, can be expressed as the product of a Lorentz transformation and a scale transformation, giving us a direction to aim for.
One may be suspcious of conformal symmetry when studying EFT as the latter has an explicit cutoff scale.
However, conformal symmetry has found utility in the study of EFT.
In particular, the Hilbert series technique for operator counting in EFTs~\cite{Jenkins:2009dy, Hanany:2010vu, Lehman:2015via, Henning:2015daa} relies on putting fields into conformal representations~\cite{Henning:2015alf, Henning:2017fpj} to account for equation of motion and integration by parts relations.
Our use of conformal symmetry is of a similar spirit in that we will be interested in representations of fields under the conformal group.
With the Hilbert series technique, the free field piece of the EFT Lagrangian is conformally invariant and the EFT power counting expansion quantifies deviations from this invariance.
Here, the piece of the Lagrangian that is invariant under operator dimension parity is also classically invariant under the center of the conformal group, $Z(Spin(4, 2)) = \Z4$.
We say classicially because operators usually develop an anomalous dimension.
Therefore, the $\Z4$ \textit{conformal center symmetry} will generally be broken once operator dimension is upgraded to scaling dimension in the quantum theory.\footnote{
	We have been implicitly assuming, particularly in our analysis of anomalies, Eqs.~\eqref{eq:anom16} and~\eqref{eq:anom}, the fractionalized operator dimension parity is an internal symmetry.
	On the other hand, the conformal center symmetry is a spacetime symmetry with other potential sources of anomalies.
}
Despite this limitation, the conformal center symmetry still provides a convenient way to get the $\odp$ charge of any irrep of the Lorentz group in a given power counting scheme.

The center of a group is the set of elements that commute with every element in the group.
The elements of a $Spin$ group include products of two gamma matrices, $\gamma_k \gamma_n$ when $k \neq n$.
The interpretation of such elements in Euclidean signature is that they are rotations by $\pi$ in the $k$-$n$ plane.
For the spinor representation of $Spin(4, 2)$, this means the generator of the center symmetry is proportional to the 6D chiral gamma matrix, $\gamma_7$, as it commutes with all gamma matrices, $\gamma_k$.
Importantly, the constant of proportionality is $\pm i$, see \textit{e.g.}~\cite{Csaki:1997aw}, which is what leads to a $\Z4$ rather than $\Z2 \times \Z2$ center.
In $3 + 1$ spacetime dimensions and Lorentzian signature, the conformal center symmetry, therefore, acts as a dilatation with scale factor $-1$, followed by a boost of $i \pi$ along the $x$ axis, finished off by a rotation by $\pi$  about that same $x$ axis.
The appearance of an imaginary rapidity may seem strange.
However, in terms of the complexified Lorentz group, this is simply a rotation by $2 \pi$ under $SU(2)_l$.
We can instead pick out $SU(2)_r$ with a boost by $-i \pi$. 
This combination of ``rotations'' generated by the conformal center symmetry has the same effect on the fields as the $\odp$ symmetry.
Thus, at least classically, the fractionalized operator dimension parity, $\odp$, coincides with the center symmetry of the conformal group.
Additionally, we see that the charge of a field can be expressed as 
\begin{equation}
\label{eq:X}
\odp \bmod 4 = 2 d - 4 j_r \bmod 4 = 4 j_l - 2 d \bmod 4 .
\end{equation}
Swapping $j_r$ and $j_l$ in~\eqref{eq:X} flips the charge assignments of the left- and right-handed fermions.
Eq.~\eqref{eq:X} is the generalization of Eq.~\eqref{eq:charges} to all irreps of the Lorentz group and is also consistent with any power counting scheme where (fermions)bosons have (half-)integer dimension.
For example, left-handed Weyl fermions have $\odp_\psi = 3$ under the mass dimension power counting scheme but instead have $\odp_\psi = 1$ when using the power counting scheme of Ref.~\cite{Buchalla:2013eza}.

%%%%%%%%%%%%%%%%%%%%%%%%%%%%%%%%%%%%%%%%%%%%%%%%%%%%%
\section{Baryon and Lepton Number Violation in $\nu$SMEFT}

For all its successes, the renormalizable Standard Model incorrectly predicts that neutrinos are massless.
The observation of neutrino oscillations~\cite{Super-Kamiokande:1998kpq, SNO:2002tuh} requires at least two neutrinos to be massive.
Insisting that the $\odp$ symmetry is free of anomalies solves this problem by requiring sterile neutrinos in the theory, which then provides a Dirac mass for the neutrinos.
It is widely understood that $(\nu)$SM is the low-energy limit of an EFT.
We now investigate what phenomenological implications there can be for $\nu$SMEFT since the $\odp$ symmetry would forbid operators of odd mass dimension.
Starting from Eq.~\eqref{eq:result}, using the definitions of baryon, $B$, and lepton, $L$, number, and imposing a constraint from hypercharge, $Y$, invariance one can show that
\begin{align}
X &= (4 + 24 p) Y + (-5 + 12 q) B + (1 + 4 r) L , \nn
\odp &\equiv X \pmod 4, 
\end{align}
for integers $p$, $q$, and $r$.
An analogous relation exists using $SU(2)_{\rm weak}$ instead of hypercharge.
A particular relation of $\odp$ in $\nu$SMEFT is to the $X$ gauge boson from Grand Unified Theories (GUTs), $SO(10) \rightarrow SU(5) \times U(1)_X$ with $X = 4 Y - 5 (B - L)$.\footnote{
	See \textit{e.g.}~\cite[\textsection~93]{Workman:2022ynf} for a review of GUTs.
}
Moving from the field level to the operator level, a gauge invariant operator will always have $\Delta Y = 0$, and  we arrive at the result of Ref.~\cite{Kobach:2016ami}
\begin{equation}
\label{eq:B-L}
\frac{\Delta B - \Delta L}{2} \equiv d \pmod 2 
\end{equation}
where $\Delta B$ and $\Delta L$ are the amounts by which an operator violates baryon and lepton numbers, respectively.

\begin{figure}
	\centering
	\includegraphics{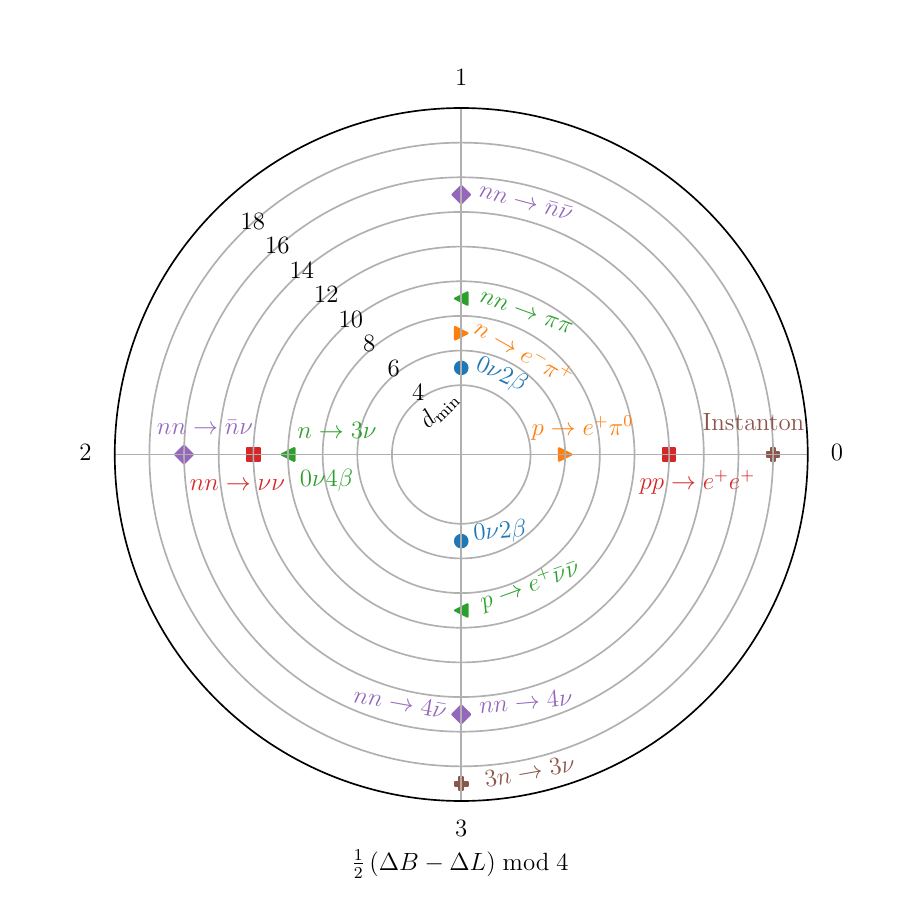}  
	\caption{
		Representative processes from Ref.~\cite{Heeck:2019kgr} that violate baryon and lepton numbers by $\Delta B$ and $\Delta L$ units, respectively.
		The radial axis is the lowest mass dimension in $\nu$SMEFT, $d_{\rm min}$, where an operator can generate the process. 
		The azimuthal axis is $\tfrac{1}{2} \left(\Delta B - \Delta L\right) \bmod 4$.
		Processes with the same color/marker combination have the same number of fermions in their $d_{\rm min}$ operators.
		Only the horizontal axis, where $\tfrac{1}{2} \left(\Delta B - \Delta L\right)$ is even, is allowed by the $\Z4^\odp$ symmetry.
	}
	\label{fig:dmin_BminusL}
\end{figure}

If the $\Z4^\odp$ symmetry is imposed, only operators in the $\nu$SMEFT with even $\tfrac{1}{2} \left(\Delta B - \Delta L\right)$ are permitted. 
This statement is true in the full quantum theory as $\nu$SMEFT is anomaly-free.
Notably, this symmetry would forbid all operators with $\Delta L = 2$ and $\Delta B = 0$, including the dimension-5 Weinberg operator, which in turn forbids Majorana neutrinos.
We wish to classify processes as to whether they are allowed or forbidden by the symmetry.
Various processes with leading-order contributions from $(\nu)$SMEFT operators were plotted on a $\Delta B$ versus $\Delta L$ grid in Ref.~\cite{Heeck:2019kgr} (see also~\cite{Helset:2019eyc}).
Figure~\ref{fig:dmin_BminusL} transforms~\cite[Fig.~1]{Heeck:2019kgr} to a coordinate system where all processes not on the horizontal axis would be forbidden by the $\odp$ symmetry, providing a visual classification tool, and emphasizing that imposing $\odp$ severely restricts the set of allowed baryon and lepton number violating processes.
One of the allowed processes, despite these restrictions, is the well-known proton decay channel, $p \to e^+ \pi^0$, which could be mediated by the aforementioned $X$ gauge boson.
The radial axis of Fig.~\ref{fig:dmin_BminusL} is the lowest mass dimension in $\nu$SMEFT, $d_{\rm min}$, where an operator can generate the process, and processes with the same color/marker combination have the same number of fermions in their $d_{\rm min}$ operators.
The azimuthal axis is $\tfrac{1}{2} \left(\Delta B - \Delta L\right) \bmod 4$, rather than mod 2, to highlight that it is still possible to violate baryon minus lepton number when $\odp$ is imposed, \textit{e.g.}~through neutrinoless \textit{quadruple} beta decay ($0\nu4\beta$).
The processes in Fig.~\ref{fig:dmin_BminusL} are not exhaustive.
For example, upcoming experiments at the European Spallation Source~\cite{Santoro:2023jly, Santoro:2023izd} would not see neutron-antineutron oscillations in this scenario. 

The presence of the $\Z4^\odp$ symmetry in SMEFT would raise questions about the nature of the UV theory that completes SMEFT.
Surprisingly, in some cases, $\Z4^\odp$ can emerge in SMEFT when it is not present in the UV theory. 
This can happen because of the peculiarity in SMEFT where baryon minus lepton number is directly related to operator dimension, see Eq.~\eqref{eq:B-L}.
SMEFT effectively has two symmetries in this scenario, and a UV theory only needs to have one of the two symmetries for SMEFT to have both.
For example, adding a real scalar singlet or real scalar electroweak triplet to the SM field content breaks $\Z4^\odp$ but respects $\Z2^{B-L}$.
It is known that when these real scalar fields are integrated out, they only yield SMEFT operators that respect $\Z4^\odp$, see e.g.~\cite{Jenkins:2013fya, Manohar:2013rga}.
There are examples involving fermions and vectors as well~\cite{deBlas:2017xtg, Li:2023cwy}.

%%%%%%%%%%%%%%%%%%%%%%%%%%%%%%%%%%%%%%%%%%%%%%%%%%%%%
\section{Conclusion}

We showed that Lorentz invariant QFTs with fermions in four spacetime dimensions have a $\Z4$ symmetry, fractionalized operator dimension parity, provided that the QFT has a basis of operators where each operator has an even canonical dimension under the power counting scheme of the QFT.
As such, a larger class of 4D QFTs can be related to 3D topological superconductors than was previously realized.
It would be interesting to find new examples of connections between particle physics QFTs and TSCs or other symmetry-protected topological states of matter.
There should also be connections to theories in 5D where the fractionalized symmetry is a spatial reflection about one axis~\cite{Tachikawa:2018njr}.
This is reminiscent of the lattice implementation of domain wall fermions~\cite{Kaplan:1992bt}.

To demonstrate the utility of the $\odp$ symmetry, we imposed it on $\nu$SMEFT, showing it provides a neutrino mass mechanism and severely restricts the allowed set of processes that violate baryon and lepton number while still allowing for $B - L$ number violation.
Another area where the $\odp$ symmetry may find applications is in cosmology because when a discrete symmetry is spontaneously broken, it leads to the production of domain walls~\cite{Zeldovich:1974uw}.
Such a $\Z4$ symmetry has already been considered in the context of inflation~\cite{Kawasaki:2023mjm}.

Finally, it is desirable to have a better understanding of the connection between the fractionalized operator dimension parity, which we have assumed is an internal symmetry, and the conformal center symmetry, which is a spacetime symmetry.
Is it merely a coincidence that they are equivalent classically?

%%%%%%%%%%%%%%%%%%%%%%%%%%%%%%%%%%%%%%%%%%%%%%%%%%%%%
\section*{Acknowledgements}

We are grateful to Andrew Kobach for initially collaborating on this project and for comments on a draft of the manuscript.
We would also like to thank Jean-Fran\c{c}ois Fortin, Julian Heeck, Ken Intriligator, and Andreas Stergiou for their helpful communications.
This work made use of $\mathtt{Matplotlib}$~\cite{Hunter:2007} and $\mathtt{pandas}$~\cite{mckinney-proc-scipy-2010, reback2020pandas} to produce Fig.~\ref{fig:dmin_BminusL}.

%%%%%%%%%%%%%%%%%%%%%%%%%%%%%%%%%%%%%%%%%%%%%%%%%%%%%
\bibliographystyle{JHEP}
\bibliography{op_dim_par_frac}

\end{document}